\documentclass[aps,prd,twocolumn,superscriptaddress,runinaddress,floatfix,preprintnumbers,showpacs]{revtex4-1}
\usepackage{amsmath}
\usepackage{amssymb}
\usepackage{amsfonts}
\usepackage{amsthm}

\newtheorem{theorem}{Theorem}[section]
\newtheorem{corollary}[theorem]{Corollary}
\newtheorem{lemma}{Lemma}[section]
\newtheorem{definition}{Definition}[section]
\newtheorem{proposition}{Proposition}[section]
\newtheorem{example}{Example}[section]

\newcommand{\eqn}[1]{ \begin{equation} #1 \end{equation} }

\usepackage{graphicx}

\begin{document}


\title{Accessing High Momentum States In Lattice QCD}

\pacs{TBD}

\author{Dale S.~Roberts}
\author{Waseem Kamleh}
\author{Derek B.~Leinweber}
\author{M.~S.~Mahbub}
\author{Benjamin J.~Menadue}
\affiliation{Special Research Centre for the Subatomic Structure of
  Matter, School of Chemistry \& Physics, University of Adelaide,
  SA 5005, Australia} 

\begin{abstract}

Two measures are defined to evaluate the coupling strength of smeared
interpolating operators to hadronic states at a variety of momenta. Of
particular interest is the extent to which strong overlap can be
obtained with individual high-momentum states.  This is vital to
exploring hadronic structure at high momentum transfers on the lattice
and addressing interesting phenomena observed experimentally.  We
consider a novel idea of altering the shape of the smeared operator to
match the Lorentz contraction of the probability distribution of the
high-momentum state, and show a reduction in the relative error of the
two-point function by employing this technique.  Our most important
finding is that the overlap of the states becomes very sharp in the
smearing parameters at high momenta and fine tuning is required to
ensure strong overlap with these states.

\end{abstract}

\maketitle

\section{Introduction}

Lattice QCD has enjoyed great success as a tool for first-principles
hadron-structure calculations. Early pion electromagnetic form factor
calculations \cite{Martinelli:1987bh,Draper:1988bp} and nucleon form
factor calculations
\cite{Martinelli:1988rr,Draper:1989pi,Leinweber:1990dv} established
the formalism and presented first results establishing the challenges
ahead for obtaining precision form factors to confront experimental
data.  Nucleon form factors continue to be an active area of research
\cite{Leinweber:2004tc,Boinepalli:2006xd,Alexandrou:2010uk,
  Alexandrou:2010hf, Alexandrou:2010cm, Aoki:2010xg, Yamazaki:2009zq,
  Syritsyn:2009mx} and a comprehensive review of recent form factor
calculations can be found in \cite{Zanotti:2008zm} and references
therein.

In practice, current lattice calculations were limited to a momentum
transfer of approximately $Q^2=3\,\mathrm{GeV}^2$ due to a challenge
of increasing statistical errors.  Recently, calculations of the
nucleon and pion form factors at $Q^2=6\,\mathrm{GeV}^2$ have been
performed using variational techniques \cite{Lin:2010fv}.  In this
paper we explore very high momentum states and propose that, with
sufficient optimisation of the smearing parameters alone, momentum
transfers of the order $Q^2=10\,\mathrm{GeV}^2$ can be accomplished in
lattice hadron structure calculations.

Smearing techniques have seen wide spread use in many applications in
lattice QCD since first being applied to fermion operators
\cite{Gusken:1989qx}. The most notable impacts can be found in
spectroscopy calculations using variational methods
\cite{Michael:1985ne,Luscher:1990ck,McNeile:2000xx,Leinweber:2004it,Mahbub:2009aa,Mahbub:2010rm}.
In spite of these successes, there has been little in the way of the
optimisation of smearing parameters for high-momenta states.  For
low-momenta states there is no real need for optimization as the
overlap of states is typically slowly varying with the smearing
parameters.  In the following we reveal that this is not the case for
high-momenta states and finely tuned optimization is very beneficial
in accessing these states on the lattice.

Isolation of the ground state at 
high-momentum is essential to removing otherwise large and 
problematic excited state contaminations. 
However,  suppression of excited states through Euclidean evolution alone 
encounters a rapid onset of statistical noise.  
 We introduce two different measures to quantify the coupling of a
smeared operator to the ground state of a proton relative to the
near-by excited states, and show how these measures determine 
the optimal smeared operator for ground state isolation early in Euclidean time.

We also introduce anisotropy into the smeared operators
in the direction of momentum in an effort to improve the coupling to
these Lorentz-contracted high-momentum states.  Our results are
complementary to the variational techniques of Ref.~\cite{Lin:2010fv}
in that the optimal set of smearings for accessing a variety of
momenta can be combined to create a correlation matrix providing an
effective basis for eigenstate isolation.

\section{Two-Point Functions}

The two-point function of a baryon on the lattice in momentum space is
given by
\eqn{G_2(\vec{p},t) = \sum_xe^{-ip\cdot x}\langle\Omega\vert\chi_i(x)\bar\chi_i(0)\vert\Omega\rangle,}
where $\chi_i$ and $\bar\chi_i$ annihilate and create the baryon
respectively at the sink point $x$ and source point $0$ and the index
$i$ admits various spin-flavor structures for the interpolators.  In
the case of the proton, the annihilation operator is
\begin{equation}
\chi_1 = \epsilon^{abc}(u_a^TC\gamma_5d_b)u_c,
\end{equation}
where $u$ and $d$ represent the spinors for the up and down quarks
respectively and $C$ is the charge conjugation matrix. It can be
shown that, for positive parity states,
\eqn{ G_2(\vec{p},t) = \sum_B 
  \frac{\gamma\cdot p + m}{2\, E_B} \, \lambda_B \, e^{-E_B\, t} \, 
\, , } 
where the sum over $B$ represents the ground and excited states of the
baryon.  It is common to average the $(1,1)$ and $(2,2)$ elements of the
Dirac matrix where the signal for positive parity states is large.  At
zero momentum, the Dirac matrix contribution is then 1.  The
coefficient $\lambda_B$ provides a measure of the total overlap of
$\bar\chi_i$ at the source and $\chi_i$ at the sink with the state
$B$.  It is the product of the source and sink overlaps which may be
different if different smearings are used at the source and the sink.
In this investigation the source will be fixed to a point source such
that variation in $\lambda_B$ is proportional to the variation in the
overlap of $\chi_i$ which will encounter a wide range of different
sink smearings.

Each state decays at a rate proportional to the exponential of its
energy.  By evolving forward in Euclidean time, excited state
contributions die away allowing the ground state to be isolated. This
is less than ideal for the calculation of three-point functions that
require effective ground state isolation close to the source to avoid large Euclidean
time evolution and an associated loss of signal.  It is for this
reason that various techniques have been implemented for earlier
Euclidean-time isolation of the ground state.

When calculating the two-point function, it is possible to choose the
momentum of the baryon. On the finite lattice, momentum
is quantised
\eqn{ \vec{p} = \frac{2\pi}{N_La}(p_x,p_y,p_z) \label{quantP}}
where $N_L$ is the spatial extent of the lattice, $a$ is the lattice
spacing and $p_x$, $p_y$, $p_z$ are integers restricted to the range
\eqn{ - \frac{N_L}{2} < p_i \le \frac{N_L}{2} \, . \label{quantPrange}}
Due to the construction of the discrete fermion propagator, momentum
input into the two-point function becomes proportional to
$\mathrm{sin}(\vec{p})$, therefore, it is only reasonable to consider
momentum states where \eqn{ |\, p_i\, | \lesssim \frac{N_L}{4}, } such
that the dispersion relation is approximately satisfied.

\section{Gaussian Smearing}

Gaussian smearing is an iterative procedure applied to the source or
sink of the two-point function in order to improve the relative
coupling to the ground state of the particle.  Consider
\eqn{ \chi_{i+1}(x) = F(x,y)\chi_i(y). }
with \cite{Gusken:1989qx}
\begin{eqnarray}
F(x,y) &=& (1-\alpha)\, \delta_{xy} \\
&&+ \frac{\alpha}{6} \sum_{\mu=1}^3 \left (
U^\dagger_\mu(x-a\hat\mu)\, \delta_{x-\hat\mu,y} + 
U_\mu(x)\, \delta_{x+\hat\mu,y} \right ) \, , \nonumber 
\end{eqnarray}
where $\alpha$ is a constant, which we set to $0.7$. We can introduce
anisotropy to the smearing by introducing a new constant $\alpha_x$,
which will act only in the $x$ direction, the expression for the
smearing then becomes,
\begin{widetext}
\eqn{ F(x,y) = (1-\alpha_o)\, \delta_{xy} + \frac{\alpha_x}{6} \left (
  U^\dagger_1(x-a\hat x)\, \delta_{x-\hat x,y} + U_1(x)\,
  \delta_{x+\hat x,y} \right ) + \frac{\alpha}{6}\sum^3_{\mu=2} \left
  (U^\dagger_\mu(x-a\hat\mu)\, \delta_{x-\hat\mu,y} + U_\mu(x)\,
  \delta_{x+\hat\mu,y} \right ) \label{anissmear} }
\end{widetext}
where $\alpha_o=0.7$ and $\alpha$ and $\alpha_x$ are normalised such that
\eqn{ \frac{4\alpha + 2\alpha_x}{6}=\alpha_o \, .}

\section{Measures}

Gusken \cite{Gusken:1989qx} introduced the measure 
\eqn{ R=\frac{G_2(t')\, e^{+m_0\, t'}}{G_2(0)} \, , }
for quantifying the ground state isolation of a hadron.  By taking a
point, $t'$, sufficiently late in time such that the excited state
contributions become negligible, the ground state can be evolved back
to the source via $e^{+m_0\, t'}$ to evaluate the fraction of $G_2(0)$ it holds.
However, with sufficient smearing, states can contribute negatively to
the two-point function, allowing this ratio to exceed $1$ and making
it difficult to interpret the results.

The first measure we introduce follows from this idea by determining
the deviation of $G_2(t)$ from the ideal two-point function of a
single ground state.
It is
similar in principle to Gusken's measure, however, it is capable of
taking into account the presence of states with negative coupling to
the operator. The measure, $M_1$ is defined as,
\eqn{ M_1 = \frac{-1}{t_f-t_i+1} \sum^{t_f}_{t=t_i} 
\frac{\left (e^{-E_0 (t-t_0)} - \tilde G_2(t) \right )^2}{\tilde
  G^2_2(t)}  \, ,
\label{M1}}
where $\tilde G(t) = G(t)/G(t_0)$. The factor $-1$ makes this measure
maximal when $G(t)$ is a pure exponential of the ground state.  The
energy $E_0$ is determined from a $4\times 4$ source-sink-smeared
variational analysis \cite{Mahbub:2009nr} of the zero momentum
state with the correct dispersion relation applied for finite-momentum
states.

Another common method of extracting coupling effectiveness is to
perform a four parameter, two exponential fit on a region close to the
source of the two-point function, {\it i.e.}
\eqn{G_{\mathrm{fit}}=a_1e^{-a_2t}+b_1e^{-b_2t}.}
However, this method tends to prove unreliable with the parameters
varying with the fit window.  The method is limited by the fact that
it can not take into account any states with higher energy than the
two considered.

The second measure we introduce works similar to this.  However, the
parameters of the exponentials are predetermined by a variational
analysis \cite{Mahbub:2009nr}.  This leads to a simple linear fit
of known exponentials, {\it i.e.}
\eqn{ G_{\mathrm{fit}}=\lambda_0\, e^{-E_0\, t} + \lambda_1\,
  e^{-E_1\, t} + \lambda_2\, e^{-E_2\, t} \, . }
We can then find the proportion of the $i$-th state in the two-point
function with the measure
\eqn{ M_{2,i} = \frac{|\, \lambda_i \, |}{\sum_k \, |\, \lambda_k \,
    |} \, . \label{M2}}

\section{Lattice Details}

Our calculations are performed on configurations of size $32^3\times
64$ with a lattice spacing of $0.0907\,\mathrm{fm}$ provided by the
PACS-CS collaboration \cite{Aoki:2008sm}. These lattices have $2+1$ 
sea quark flavours generated with the Iwasaki gauge action 
\cite{Iwasaki:1983ck} and the non-perturbatively improved
Clover fermion action \cite{Sheikholeslami:1985ij} with the $\kappa$ values for the
light quarks and the strange quark given by $0.13754$ and $0.13640$
respectively, and $C_{SW}=1.715$. This gives a pion mass of
$m_\pi=389\,\mathrm{MeV}$.

In order to eliminate any bias caused by smearing in the source, we use
a single set of propagators generated with a point source. All of the
smearing is then applied to the sink, making the two-point functions smearing
dependent. All momentum will be in the $x$ direction, {\it i.e.}
$p_y=0$ and $p_z=0$ in Eq.~(\ref{quantP}).

We use a $4\times 4$ correlation matrix to extract our excited state
masses, constructed from the $\chi_1$ operator with 16, 35, 100 and
200 sweeps of smearing.  We choose to use the larger basis in order to
ensure that the first three eigenstate energies are accurately
determined.

We have verified that no multi-particle states are present in the
variational analysis by applying the single-particle dispersion
relation to the zero momentum effective state masses to successfully
predict the effective masses of the same states with non-zero
momentum.  

Our error analysis is performed with the second-order
single-elimination jackknife method.  Linear fits are performed
using the normal equations with exact matrix inversion where possible
and singular value decomposition otherwise.

\section{Results}

\subsection{Isotropic Smearing}

\begin{figure}
  \begin{center}
    \includegraphics[width=0.7\linewidth, angle=90]{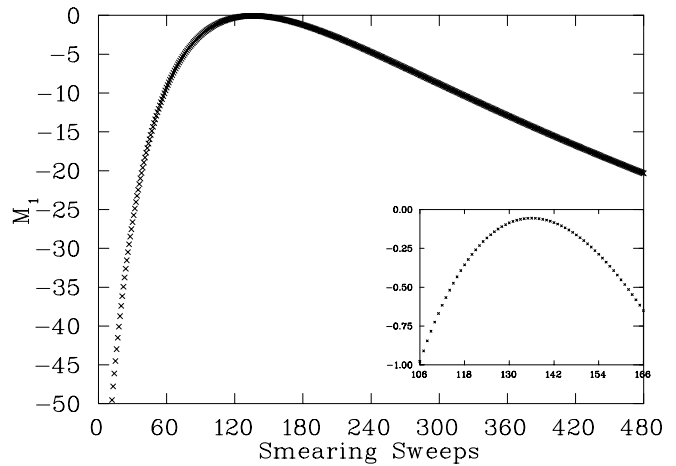}
  \end{center}
  \caption{The measure from Eq.~(\ref{M1}) at $p_x=0$ in
    Eq.~(\ref{quantP}). Deviation from the ideal two-point function
    increases by a factor of 10 less than $30$ sweeps from the
    ideal smearing level, as shown in the inset graph }
  \label{diffMeasip00}
\end{figure}

We first calculate the measure from Eq.~(\ref{M1}) where the two-point functions have been normalised $1$ time
slice after the source, with $t_i=1$ and $t_f=6$. The two-point function is calculated at every sweep of sink
smearing between $1$ and $480$, up to an rms radius of $13.68$ in
lattice units. For this particular
ensemble, the two-point function that shows the highest proportion of
ground state has $136$ sweeps of smearing at the sink, or an rms
radius of $6.92$ lattice units as seen in Fig.~\ref{diffMeasip00}. Also 
apparent is that the effectiveness 
of the smearing at isolating the ground state is significantly reduced fairly close to the optimal amount of smearing.  
At only $30$ sweeps away from the ideal number of sweeps, the deviation from the ideal two-point 
function has increased by a factor of $10$.

\begin{figure*}
  \begin{center}
    \includegraphics[width=0.32\linewidth, angle=90]{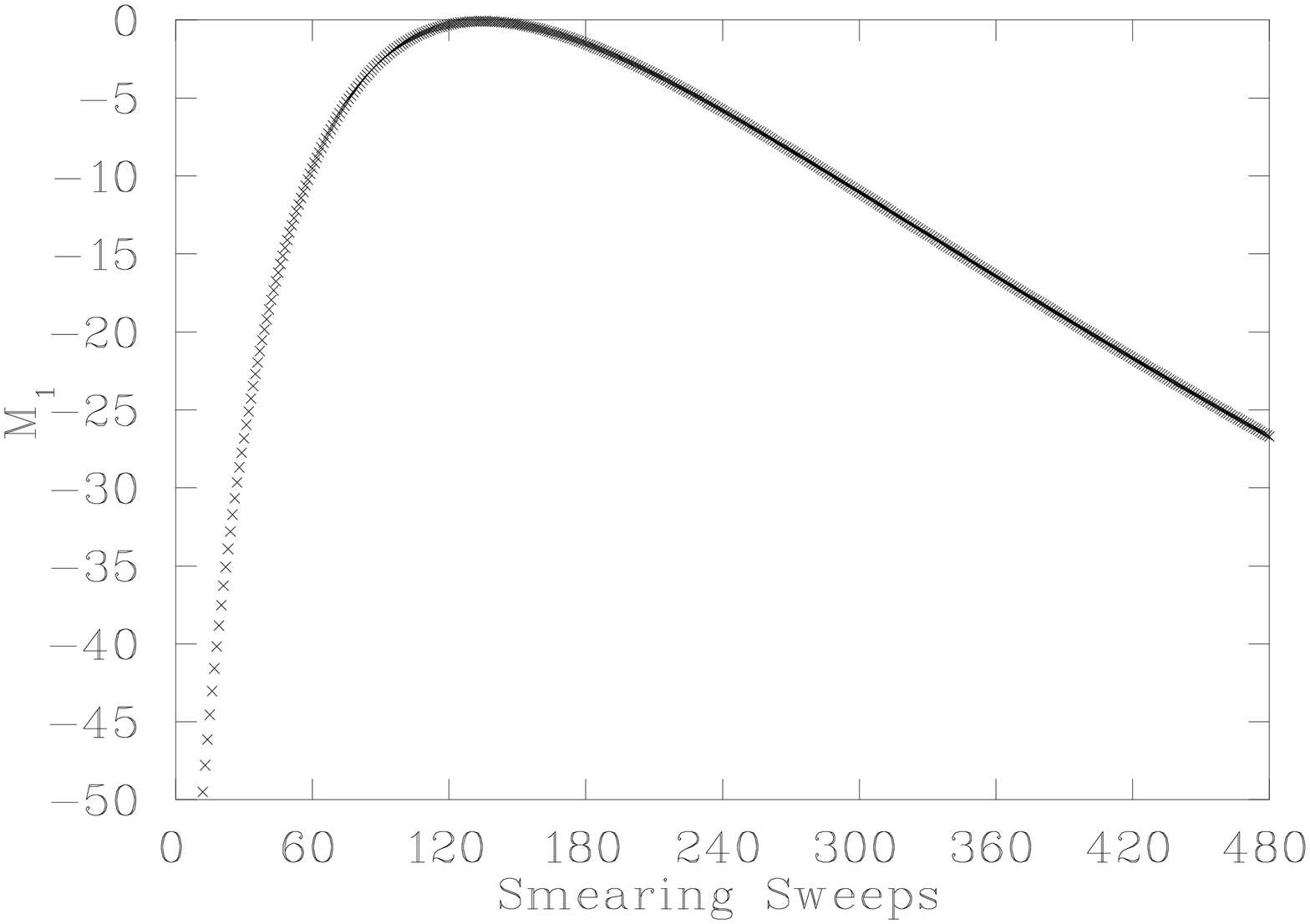} \hspace{1cm}
    \includegraphics[width=0.32\linewidth, angle=90]{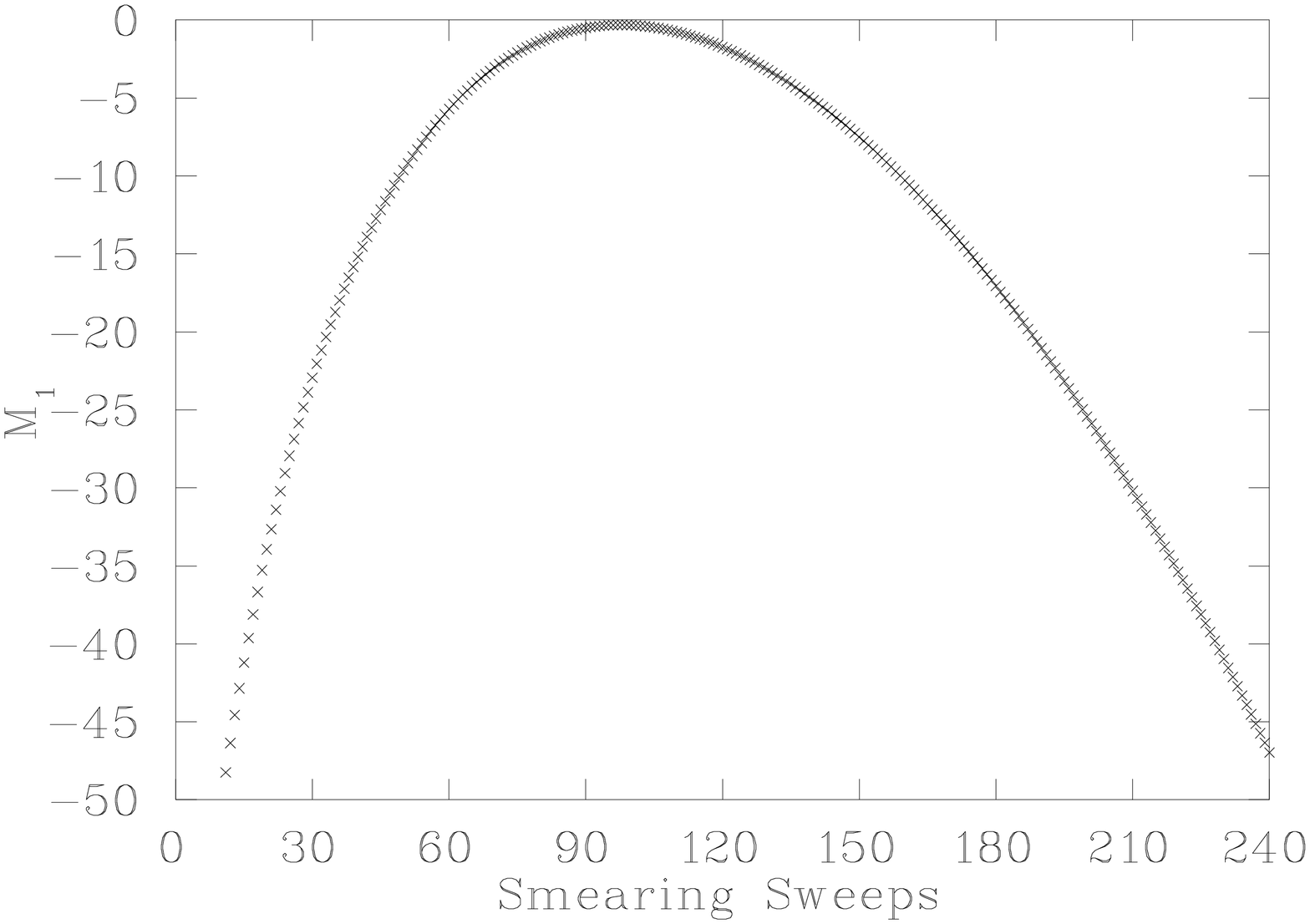}
  \end{center}
  \caption{The measure from Eq.~(\ref{M1}) at $p_x=1$ (left) and
    $p_x=3$ (right) in Eq.~(\ref{quantP}). There is little difference
    between the measure at $p_x=0$ and $p_x=1$, due to the fact that
    the probability distributions between the two momentum states are
    nearly identical.  At $p_x=3$, the rms radius of the optimal
    smearing level is smaller by a factor of $0.85$ relative to the
    $p_x=0$ state, whereas the relativistic $\gamma$ factor provides a
    Lorentz contraction factor of $\gamma^{-1} = 0.72$}.
  \label{diffMeasip01}
\end{figure*}

When we move to $p_x=1$ in Eq.~(\ref{quantP}), which gives momentum in
the $x$ direction of $427\,\mathrm{MeV}$, the ideal number of smearing
sweeps reduces by just one sweep to $135$ (rms radius $6.90$ lattice
units), as shown in Fig.~\ref{diffMeasip01}. This can be explained by
considering the relativistic $\gamma$ factor, which is given by the
ratio of the relativistic energy momentum relation and the ground
state mass. The fitted ground state mass for the proton is
$M_P=1.273(21)\,\mathrm{GeV}$, giving a relativistic energy of
$E_P\vert_{p=1}=1.343(23)\,\mathrm{GeV}$ and $\gamma=1.05$. Given that
all of the excited states are more massive, and therefore exhibit less
Lorentz contraction than the ground state, it is feasible that there
is very little difference in the probability distribution between this
state and the zero momentum state, thus the ideal amount of smearing
should be very similar to the zero momentum state.

\begin{figure*}
  \begin{center}
    \includegraphics[width=0.32\linewidth, angle=90]{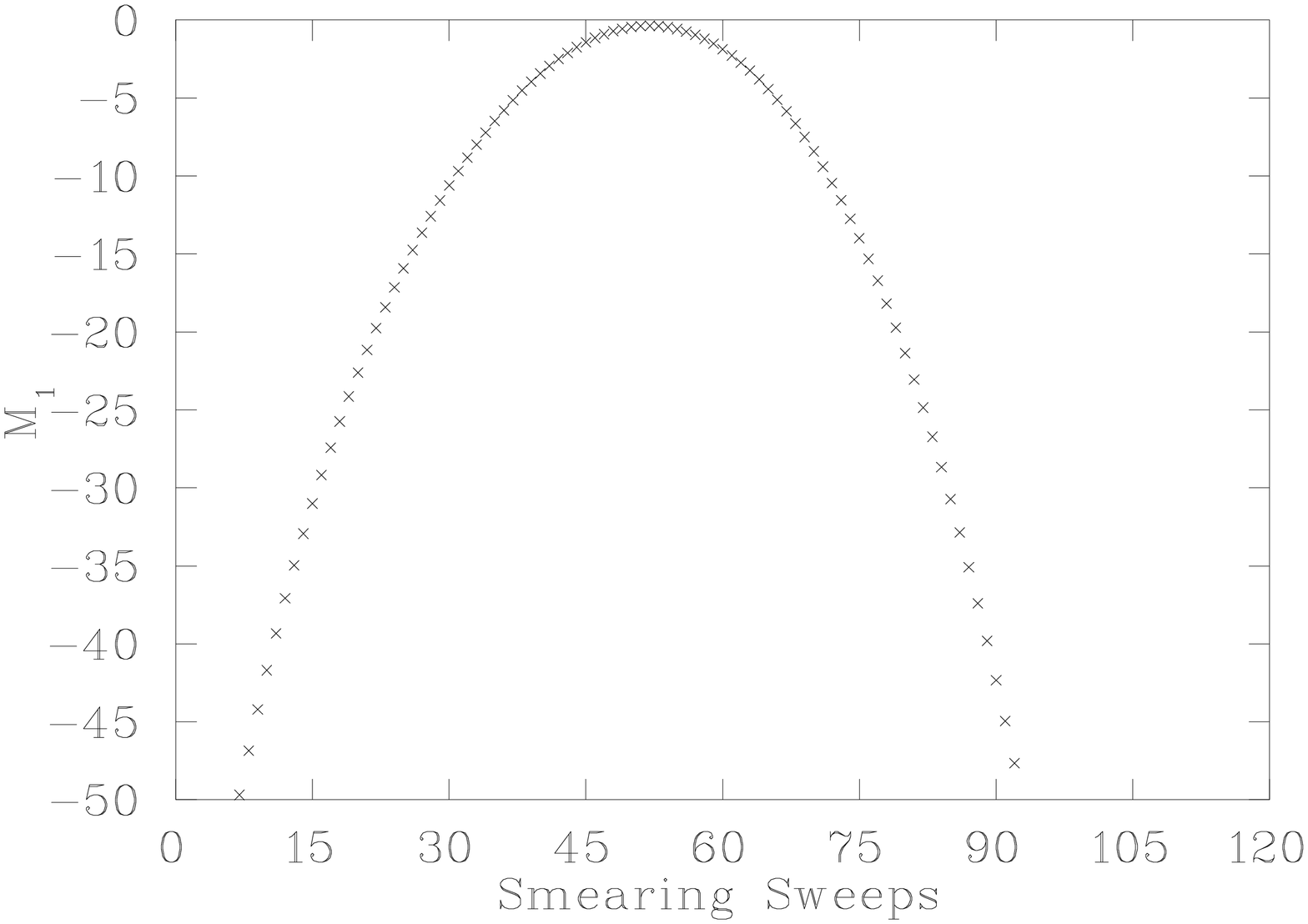} \hspace{1cm}
    \includegraphics[width=0.32\linewidth, angle=90]{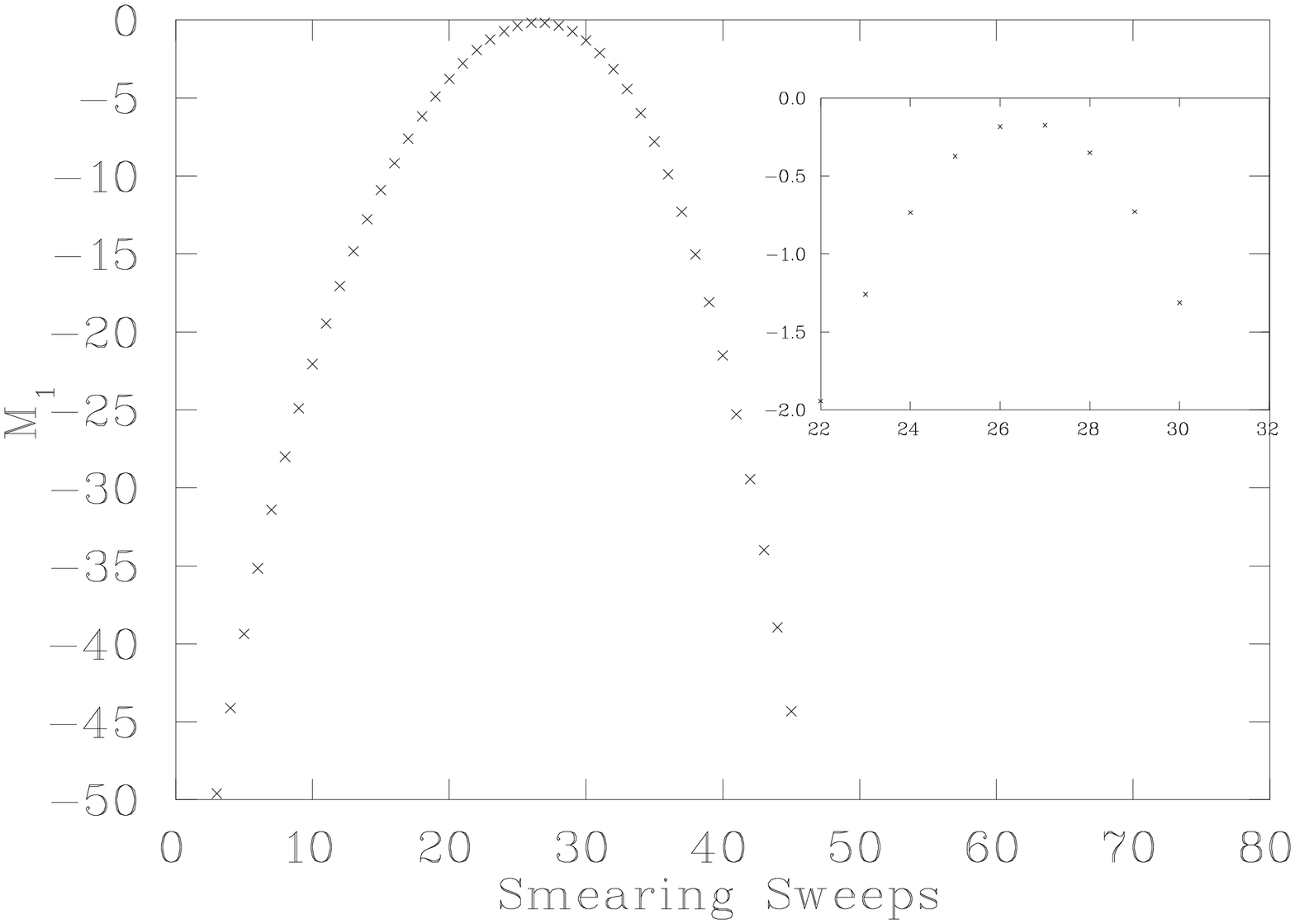}
  \caption{The measure from Eq.~(\ref{M1}) at $p_x=5$ (left) and $p_x=7$ (right) in
    Eq.~(\ref{quantP}). The value of the measure at the optimum number of smearing sweeps for this
    momentum state is approximately equal to that of the $p_x=3$
    state, indicating that good ground state isolation is possible even at higher
    momenta. At $p_x=7$, the deviation from the ideal two-point function has
    increased by a factor of 10 only $5$ sweeps from the optimal smearing
    level, as shown in the inset graph.}
  \label{diffMeasip05}
\end{center}
\end{figure*}

At $p_x=3$ in Fig.~\ref{diffMeasip01}, the optimal number of smearing
sweeps has decreased to $98$. The maximum value of the measure has
also decreased relative to the lower momentum states, indicating
relatively more excited state contamination, though still achieving
good isolation.  The ratio of the rms radius of the optimal smearing
for this state to the optimal smearing for the ground state is $0.85$,
compared to the relativistic $\gamma^{-1}$ factor of $0.72$. At $p_x=5$,
corresponding to a momentum transfer of approximately
$4.55\,\mathrm{GeV}^2$, shown in Fig.~\ref{diffMeasip05}, the optimal
number of sweeps is $52$ (rms radius $4.27$ lattice units). However,
the maximum value of the measure is close to the maximum value for the
$p_x=3$ case, indicating that very efficient isolation is possible,
even at larger momentum transfers.

Moving to $p_x=7$, equivalent to a momentum transfer of
$8.93\,\mathrm{GeV}^2$, there is significant noise far from the source
in the two-point function, even for highly optimised smearing values.
Hence we consider $t_f=5$ in the measure from Eq.~(\ref{M1}) at this
value of momentum. The ideal number of sweeps decreases to $27$
sweeps, or $3.08$ lattice units rms radius, seen in
Fig.~\ref{diffMeasip05}. Notably, the
deviation from the ideal two-point function increases by a factor of
$10$ only $5$ sweeps from this optimal value, corresponding to a
change in rms radius of less than $0.3$ lattice units.

\begin{figure}
  \begin{center}
    \includegraphics[width=0.7\linewidth, angle=90]{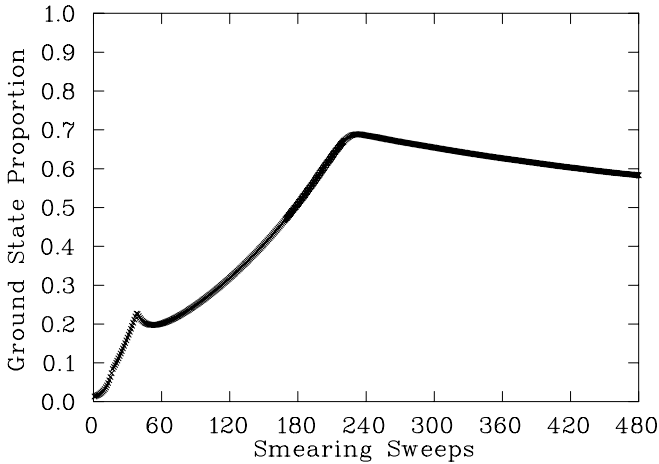}
  \end{center}
  \caption{Ground state proportion from the three exponential fit at
    $p_x=0$ in Eq.~(\ref{quantP}). There is insufficient information
    on the second excited state close to the optimal amount of
    smearing, thus requiring use of the two exponential fit to
    determine the optimal amount of smearing with this measure.}
  \label{ThreeExpGSPip00}
\end{figure}

\begin{figure}
  \begin{center}
    \includegraphics[width=0.7\linewidth, angle=90]{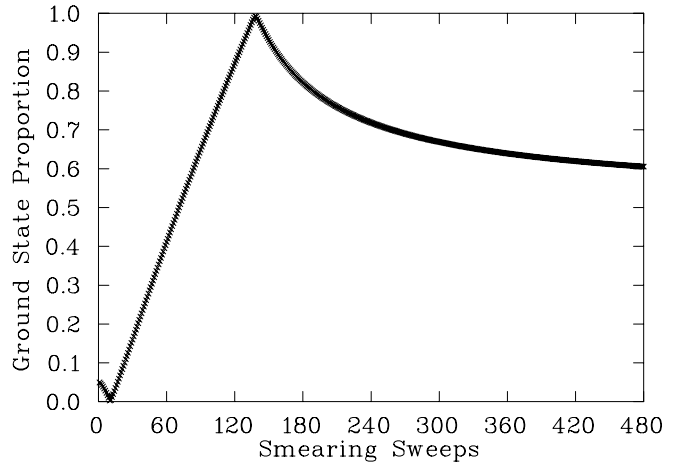}
  \end{center}
  \caption{Ground state proportion at $p_x=0$ in
    Eq.~(\ref{quantP}). Contamination due to excited states increases
    rapidly away from the optimal smearing level. There is good
    agreement between the two exponential fit here and the three
    exponential fit in Fig.~\ref{ThreeExpGSPip00} away from the
    optimum smearing levels.}
  \label{GSPip00}
\end{figure}

Using the measure described in Eq.~(\ref{M2}), we first consider the
three exponential fit between time slices $1$ and $6$ after the source
with masses $1.273(21)\,\mathrm{GeV}$, $2.301(28)\,\mathrm{GeV}$ and
$2.786(95)\,\mathrm{GeV}$ as determined in our correlation matrix
analysis. From the results in Fig.~\ref{ThreeExpGSPip00}, we can see
that, in the region where the first measure predicts ideal smearing
levels, there is a sharp change in the structure of the graph. In
order to determine the cause of this, we compare with the fits
containing only the ground and first excited
states. Fig.~\ref{GSPip00} shows that the optimal number of smearing
sweeps lies close to the value predicted by the first measure. The
overlap at the optimal number of sweeps, $138$ in this case, is
$99.31(8)\%$, indicating that, in the three exponential fit, we are
attempting to fit two quickly decaying exponentials using only
$0.69\%$ of the signal available. This leads us to believe that, in
the regions of ground state dominance where we are most interested,
the coefficient from the quickly decaying third state cannot be
determined accurately, therefore dominates well beyond where it should
be allowed to contribute at all. For this reason, we will only
consider fits using the ground and first excited states.

The contamination due to excited states in the two exponential fit at zero momentum
increases rapidly away from the optimum smearing level. Of the
smearing sweeps used to extract the masses from the variational
analysis, the one that shows the most overlap with the ground state is
$200$ sweeps, or an rms radius of $8.55$ lattice units, with
$77.69(7)\%$, or $32$ times more excited state contamination than the
optimal smearing level.

\begin{figure}
  \begin{center}
    \includegraphics[width=0.7\linewidth, angle=90]{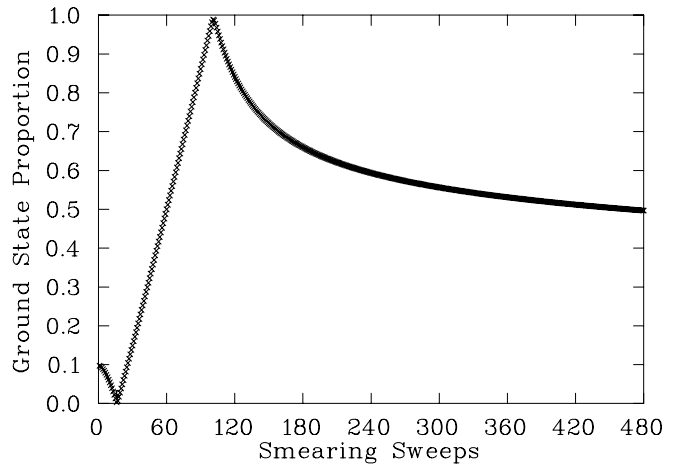}
  \end{center}
  \caption{Ground state proportion at $p_x=3$ in
    Eq.~(\ref{quantP}). As momentum increases, the contamination due
    to excited states increases more rapidly away from the ideal
    smearing level.}
  \label{GSPip03}
\end{figure}

At the first non-zero momentum state, the results present similarly to
the first measure, the optimal amount of smearing is $1$ sweep less
than that of the non-zero momentum ground state, and $2$ sweeps more than
the optimal amount determined by the first measure. At $p_x=3$ in
Eq.~(\ref{quantP}) shown in Fig.~\ref{GSPip03}, the overlap is maximised at $101$
sweeps of smearing, or an rms radius of $5.95$ lattice units, once
again agreeing within only a few sweeps of the optimum level suggested
by the first measure. Remarkably, considering the use of a point
source, the proportion of ground state present at this optimal amount of smearing is
$98.87(12)\%$.

\begin{figure*}
  \begin{center}
    \includegraphics[width=0.32\linewidth, angle=90]{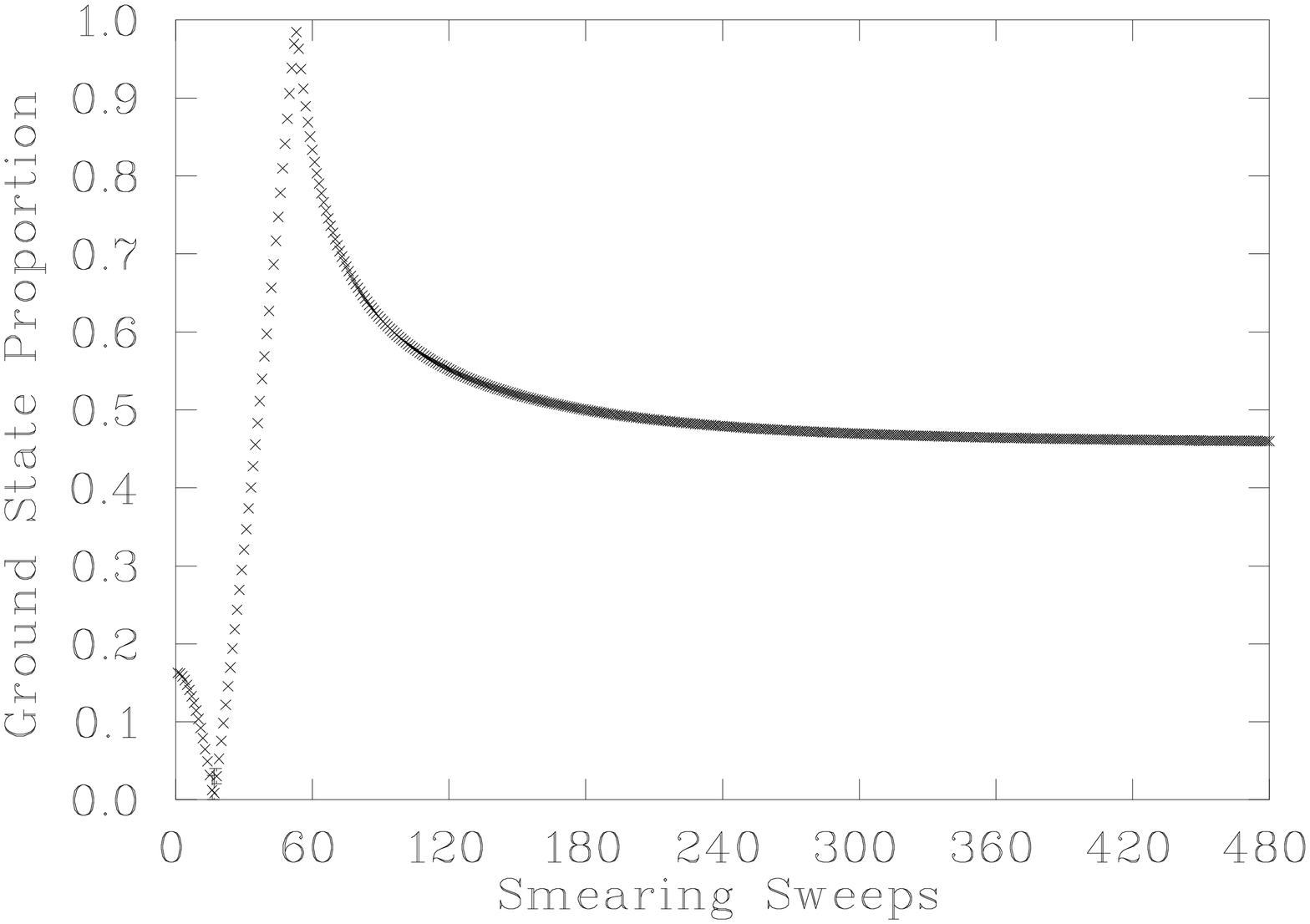} \hspace{1cm}
    \includegraphics[width=0.32\linewidth, angle=90]{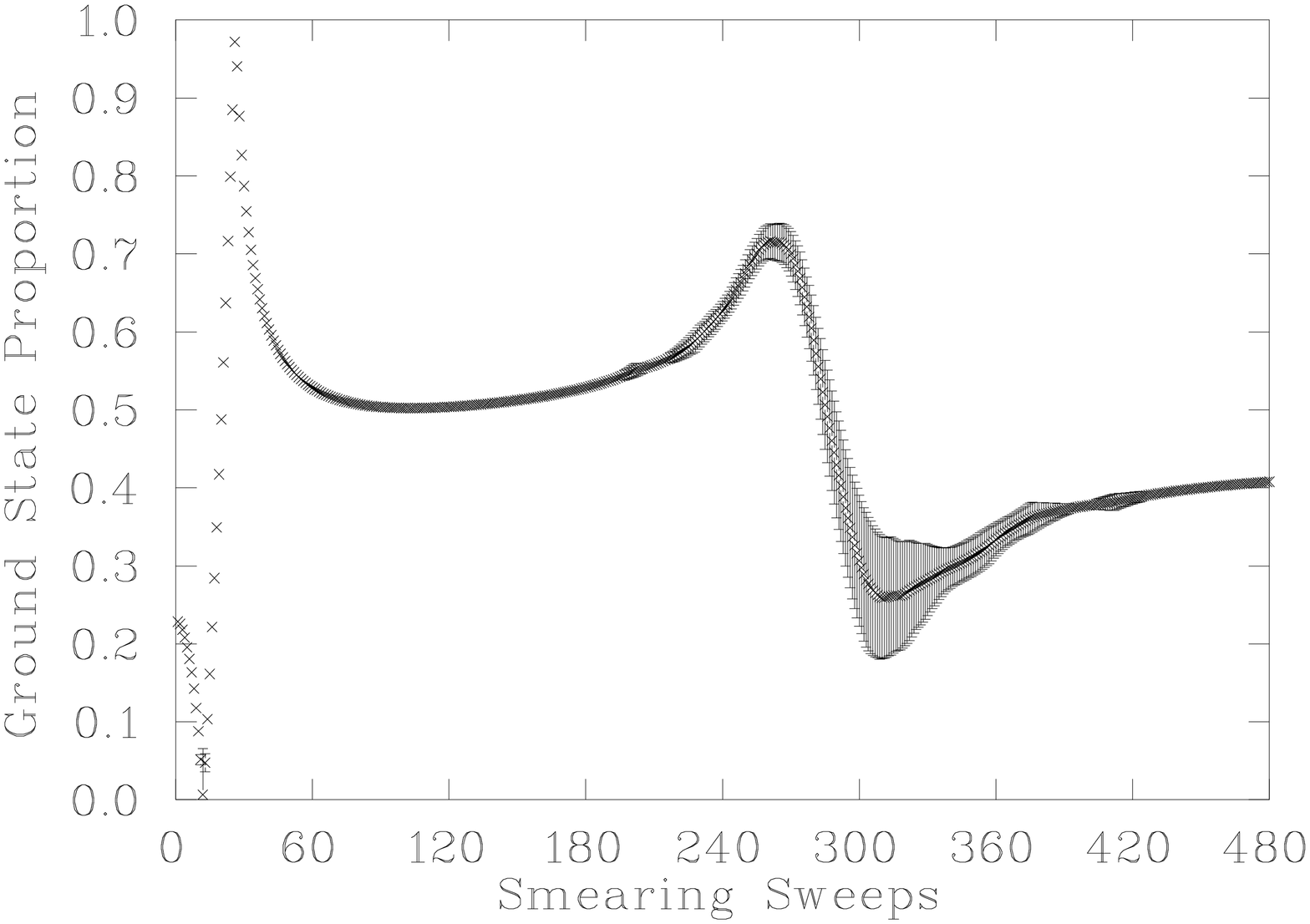}
  \end{center}
  \caption{Ground state proportion at $p_x=5$ (left) and $p_x=7$ (right) in Eq.~(\ref{quantP})
    Even at these very high momentum transfers, good overlap
    with the ground state is achieved for an optimised sink. Far from the optimal number of smearing
    sweeps at $p_x=7$, it is clear that the measure is no longer applicable, as
    there would be little, if any highly Lorentz contracted ground state present. }
  \label{GSPip05}
\end{figure*}

At $p_x=5$ and $p_x=7$ in Fig.~\ref{GSPip05} there is again good
agreement between the two measures, with the optimal smearing level
being $53$ and $26$ sweeps respectively. Even at a momentum transfer
of $8.93\,\mathrm{GeV}^2$, $97.20(20)\%$ overlap is achieved with the
ground state, and once again, very few sweeps from the optimum level,
the overlap drops dramatically. At $p_x=7$, far from the optimal
number of smearing sweeps, it is unlikely that any highly Lorentz
contracted state would couple to such a large sink. The second peak in
Fig.~\ref{GSPip05} can therefore be considered to signify a limit to
the domain of validity of the measure.

\subsection{Anisotropic Smearing}

\begin{figure*}[tph]
  \begin{center}
    \includegraphics[width=0.49\linewidth]{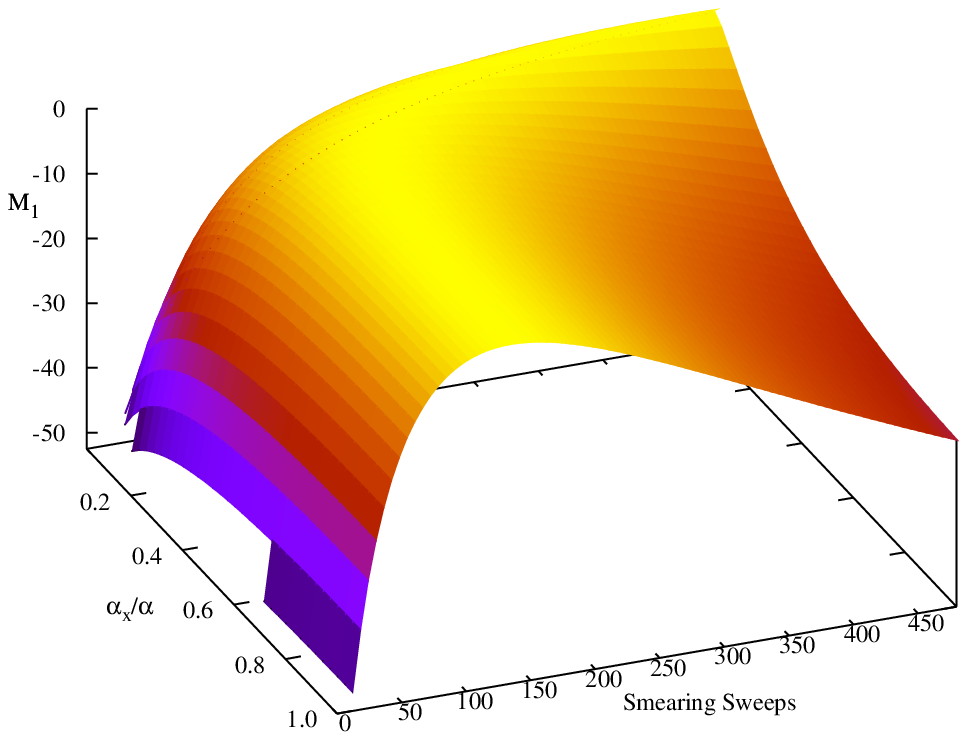}
    \includegraphics[width=0.49\linewidth]{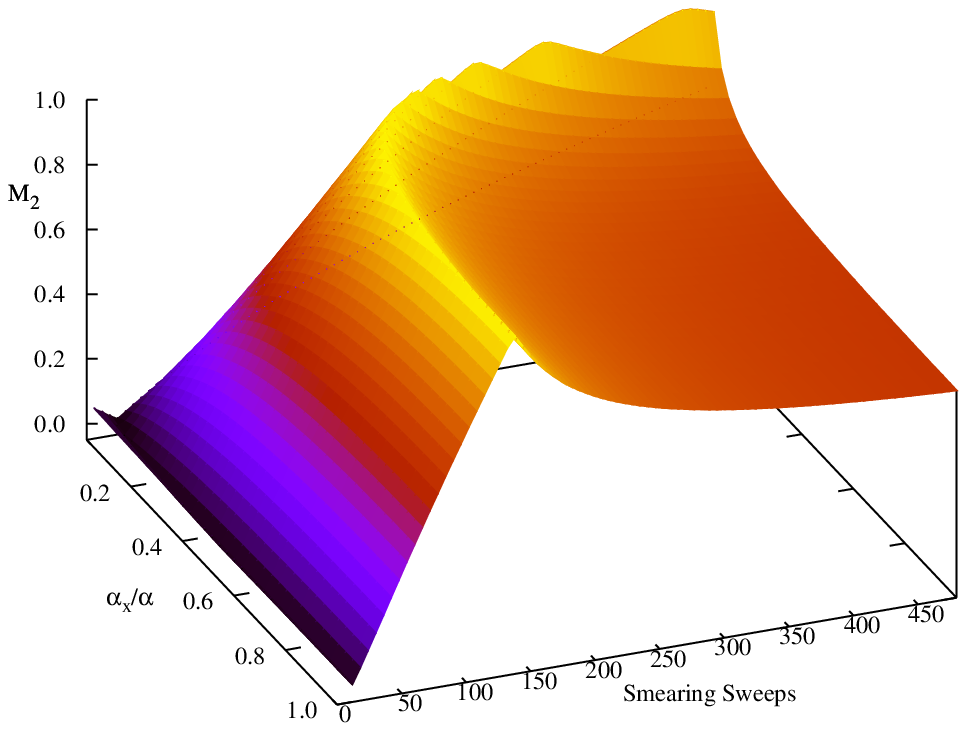}
    \vspace{-1cm}
  \end{center}
  \caption{The first measure from Eq.~(\ref{M1}) (left) and the Ground
    State Proportion (right) with anisotropic smearing at $p_x=1$ from
    Eq.~(\ref{quantP}).  Introducing anisotropy to the smearing does
    not improve the isolation of this state.  However, the Lorentz
    contraction is small so little improvement would be expected. }
  \label{M1Surfip01}
\end{figure*}

As anisotropy is introduced to the smearing as described in
Eq.~(\ref{anissmear}), we consider the first measure from
Eq.~(\ref{M1}) at the first non-zero momentum state and find that
there is no improvement to the ground state isolation, as shown in
Fig.~\ref{M1Surfip01}. There is, however, an ideal number of sweeps
that increases for decreasing $\alpha_x$ that shows approximately
equal ground-state proportion relative to the isotropic smearing case.

\begin{figure*}[tph]
  \begin{center}
    \includegraphics[width=0.49\linewidth]{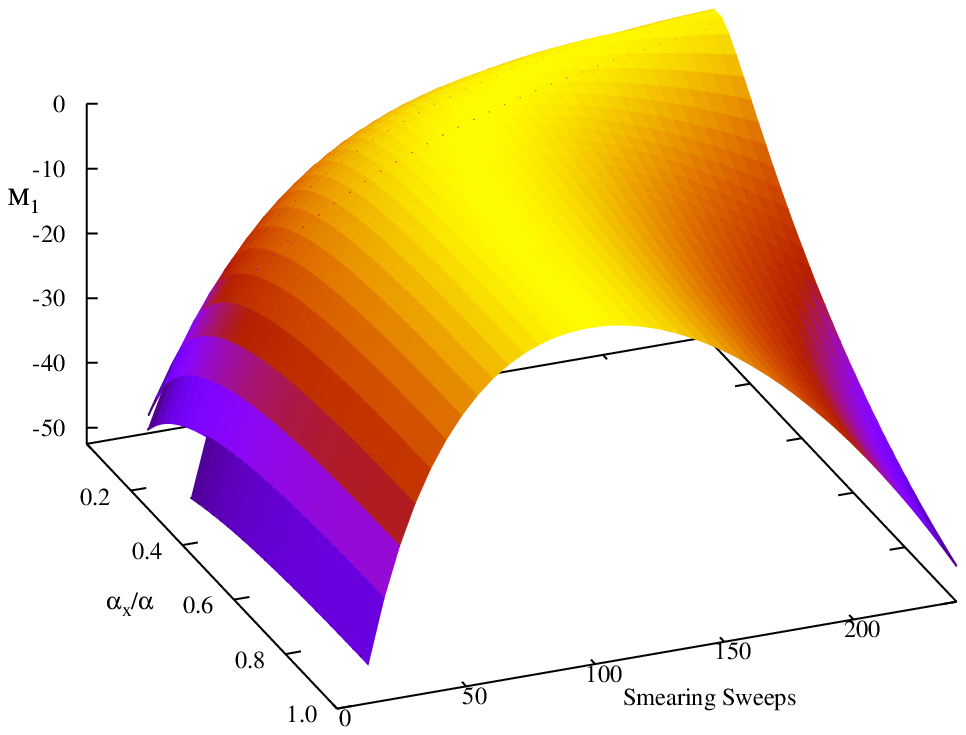}
    \includegraphics[width=0.49\linewidth]{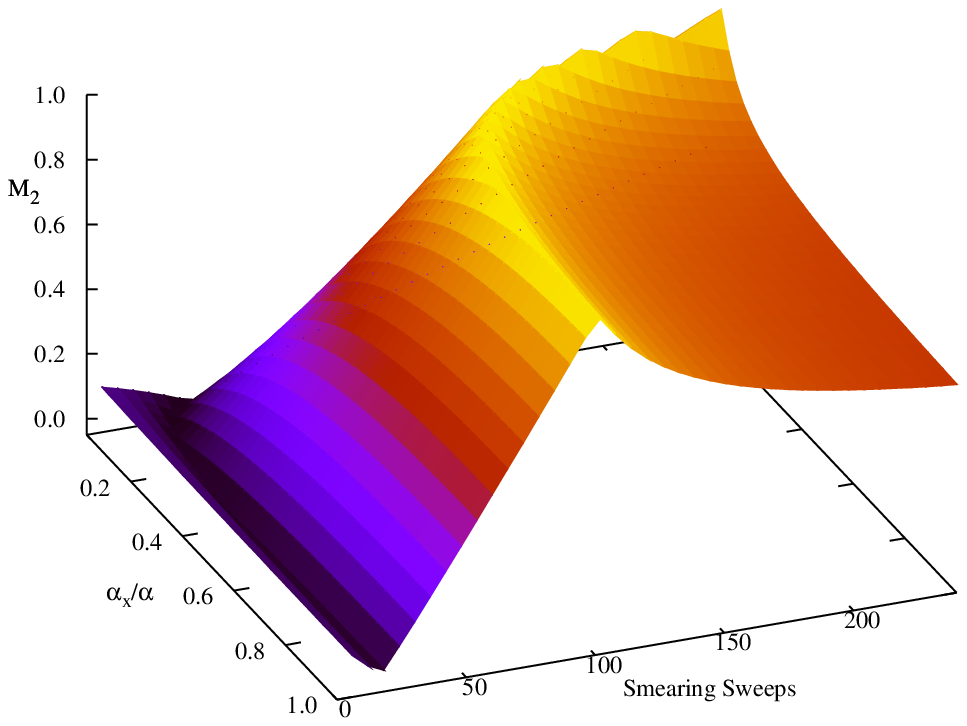}
    \vspace{-1cm}
  \end{center}
  \caption{The first measure from Eq.~(\ref{M1}) (left) and the Ground
    State Proportion (right) with anisotropic smearing at $p_x=3$ from
    Eq.~(\ref{quantP}).  No improvement is seen in the isolation of
    the ground state, in spite of the relativistic $\gamma$ factor of
    $1.39$ giving a length contraction factor of 0.72 in the $x$
    direction.}
  \label{M1Surfip03}
\end{figure*}

At $p_x=3$ in Eq.~(\ref{quantP}), in spite of the clear difference in
the smearing sweeps required to maximise overlap with the source,
Fig.~\ref{M1Surfip03} shows that introducing anisotropy to the
smearing does not result in improved isolation of the ground
state. The structure of the curve is similar to that of the $p_x=1$
state, where there is an optimal number of sweeps for every value of
$\alpha_x$ which increases with decreasing $\alpha_x$.

\begin{figure*}[tph]
  \begin{center}
    \includegraphics[width=0.49\linewidth]{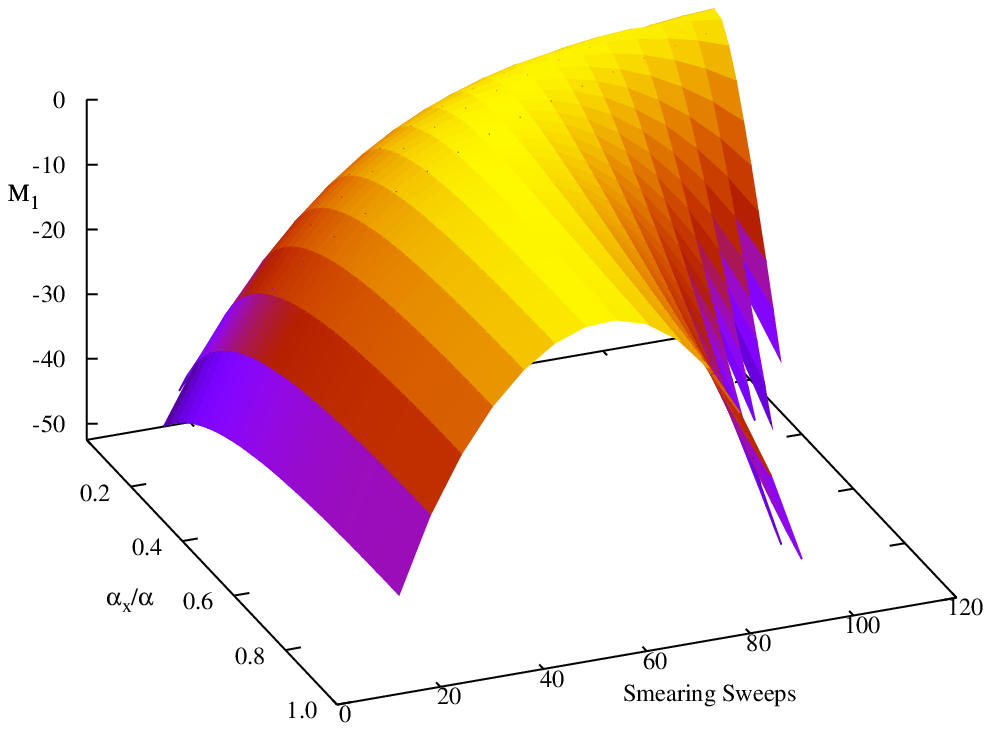}
    \includegraphics[width=0.49\linewidth]{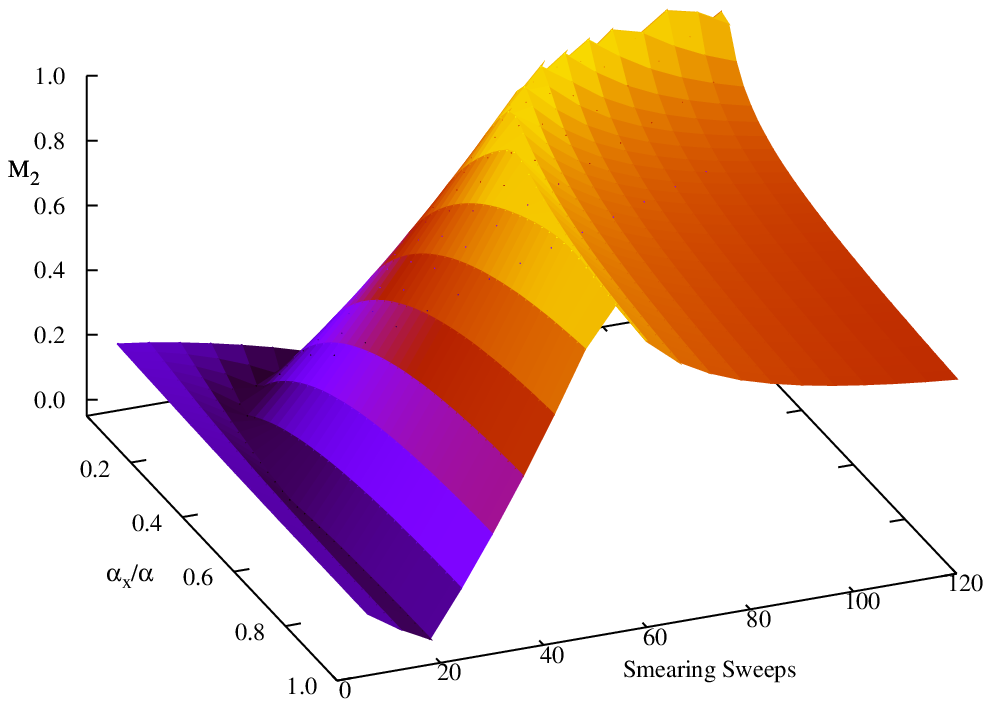}
    \vspace{-1cm}
  \end{center}
  \caption{The first measure from Eq.~(\ref{M1}) (left) and the Ground
    State Proportion (right) with anisotropic smearing at $p_x=5$ from
    Eq.~(\ref{quantP}).  The structure observed in the plots of the
    $p_x=3$ state is retained, with more sweeps of smearing required
    as anisotropy is increased.}
  \label{M1Surfip05}
\end{figure*}

\begin{figure*}[tph]
  \begin{center}
    \includegraphics[width=0.49\linewidth]{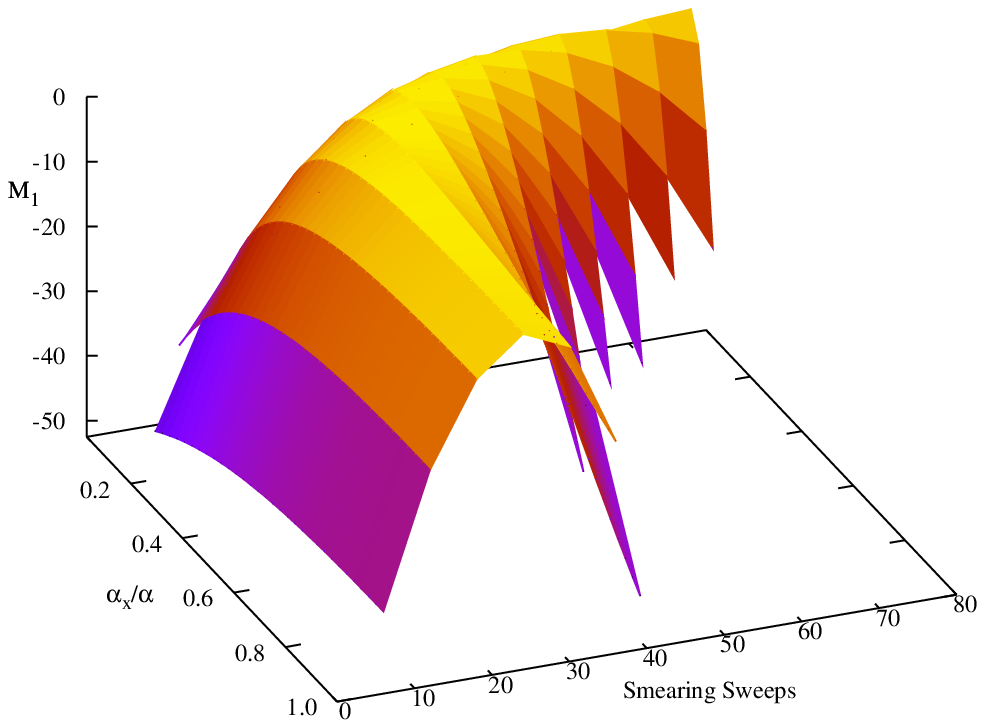}
    \includegraphics[width=0.49\linewidth]{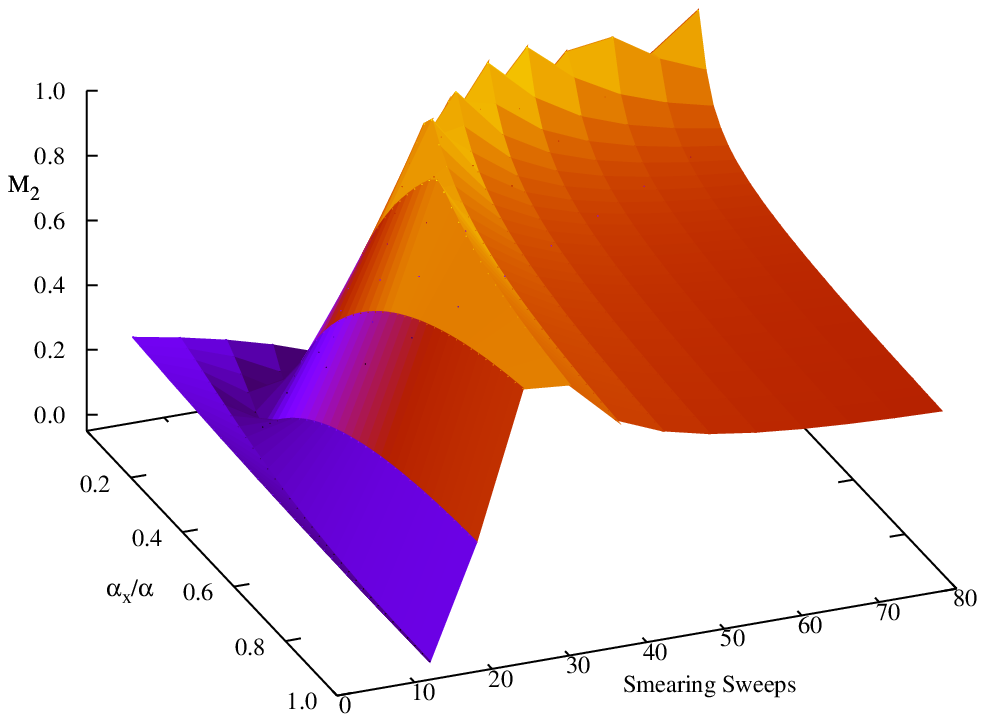}
    \vspace{-1cm}
  \end{center}
  \caption{The first measure from Eq.~(\ref{M1}) (left) and the Ground
    State Proportion (right) with anisotropic smearing at $p_x=7$ from
    Eq.~(\ref{quantP}). Even at a momentum of $2.99\,\mathrm{GeV}$,
    anisotropy in the smearing does not improve isolation of the
    ground state.}
  \label{M1Surfip07}
\end{figure*}

Once again, there is no improvement in the ability of anisotropic
smearing to isolate the ground state at the momentum of $p_x=5$, as
shown in Fig.~\ref{M1Surfip05}. The structure revealed in the lower
momentum states persists for this state and for the $p_x=7$ state in
Fig.~\ref{M1Surfip07}.  From these results, optimisation of the number
of smearing sweeps alone is sufficient to achieve good isolation of
the ground state of the two-point function at a range of momenta.

We now investigate how anisotropic smearing affects the
signal-to-noise ratio or quality of the two-point function at high
momenta.  Since we have ensured that the ground state is isolated as
close to the source as possible, we now determine the quality of the
signal a few time slices away from the source.  We consider the
relative error of the two-point function four times slices after the
source at the optimal number of smearing sweeps for each value of our
anisotropy parameter, $\alpha_x$.

\begin{figure}
  \begin{center}
    \includegraphics[width=0.7\linewidth, angle=90]{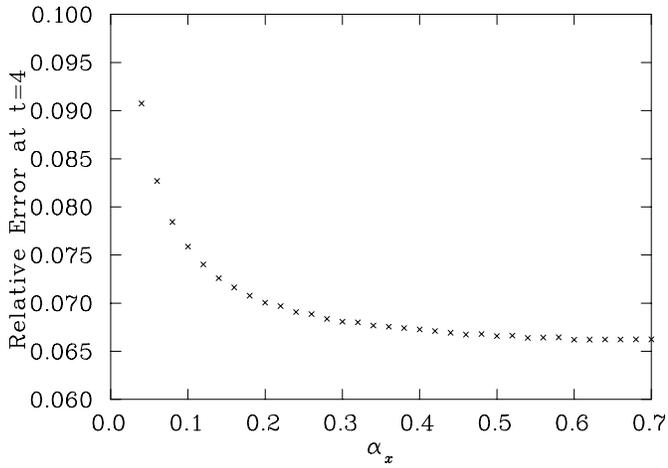}
  \end{center}
  \caption{Relative error in the two-point function measured four time
    slices after the source for $p_x=3$ as in Eq.~(\ref{quantP}).  At
    this momentum, isotropic smearing provides the best relative
    error. }
  \label{relerrip03}
\end{figure}

\begin{figure*}
  \begin{center}
    \includegraphics[width=0.32\linewidth, angle=90]{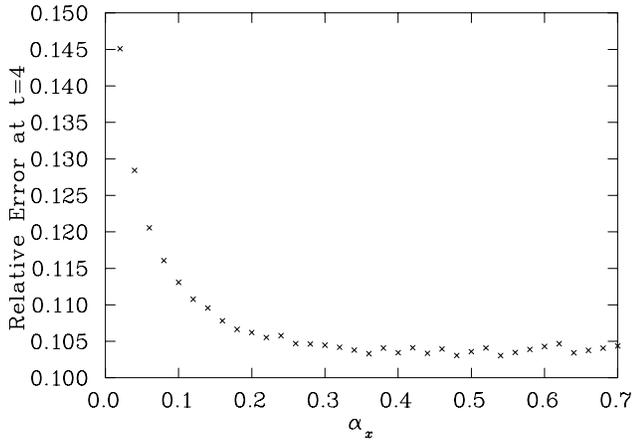} \hspace{1cm}
    \includegraphics[width=0.32\linewidth, angle=90]{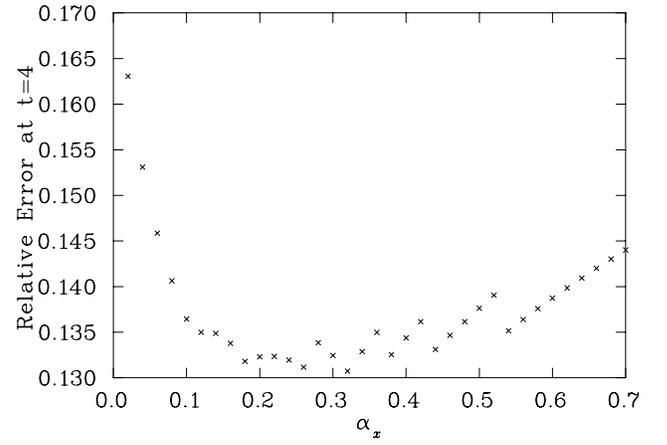}
  \end{center}
  \caption{Relative error in the two-point function measured four time
    slices after the source for $p_x=5$ (left) and $p_x=7$ (right) as
    in Eq.~(\ref{quantP}).  At $p_x=5$, there is a small amount of
    improvement for anisotropic smearing at $\alpha_x/\alpha$ in the
    region of $\gamma^{-1} = 0.51$.  At $p_x=7$, a 10\% improvement in
    the relative error is seen for values of $\alpha_x \simeq 0.26$ to
    0.32 where $\alpha_x/\alpha = 0. 37$ to 0.46, in accord with the
    value of $\gamma^{-1} = 0.39$ predicted by Lorentz contraction.
    Note that the emergent banding structure reflects a change in the
    optimal number of smearing sweeps by one.}
  \label{relerrip05}
\end{figure*}

For $p_x=3$, Fig.~\ref{relerrip03} shows the two-point function at
$t=4$.  The smallest relative error occurs when the smearing is
isotropic.  Increasing the momentum to $p_x = 5$ lattice units shows
that there is only a small improvement to the relative error for
values of $\alpha_x \sim 0.48$.  It is worth noting that the first of
the minima visible in Fig.~\ref{relerrip05} at $\alpha_x = 0.36$
corresponds to the anisotropy expected due to Lorentz contraction as
$\alpha_x/\alpha = 0.51$ equals $\gamma^{-1} = 0.51$.

The banding structure visible in Fig.~\ref{relerrip05} is a result of
the optimal number of smearing sweeps increasing for decreasing values
of $\alpha_x$.  Each discontinuity in the graph for $\alpha_x > 0.36$
is the result of the optimal number of smearing sweeps decreasing by
$1$. It is an artifact resulting from the density of the points in
$\alpha_x$ being much finer than the density of the points in the
number of smearing sweeps.

Moving to $p_x=7$ in Fig.~\ref{relerrip05} we see a distinct
improvement in the correlation-function relative error when anisotropy
is introduced.  Both $\alpha_x=0.26$ and $0.32$ provide a $10\%$
reduction in the error relative to that observed at the isotropic
value of 0.7.  The values of $\alpha_x \simeq 0.26$ to 0.32 provide
$\alpha_x/\alpha = 0. 37$ to 0.46, in accord with the value of
$\gamma^{-1} = 0.39$ predicted by Lorentz contraction.

\section{Conclusion}

We have presented two new measures of the effectiveness of smeared
operators in isolating the ground state of a hadron in the two-point
function.  Both measures show good agreement with each other. We have
performed a detailed analysis of ground state isolation with each
measure and have shown that optimisation of the smearing can lead to
remarkable improvement to the ground state isolation.  Furthermore,
the ability to isolate the ground state decreases dramatically a few
sweeps from the optimal number of smearing sweeps for the higher
momentum states.  In selecting a basis for a correlation matrix
analysis, these optimal smearing parameters are preferred.

On the introduction of anisotropy to the smearing, we found that there
was no appreciable improvement to the overlap with the ground state.
The relative proportion of the ground state for an isotropic source is
already high.  Optimising the number of sweeps of isotropic smearing
alone is sufficient to ensure maximal isolation of high-momentum
ground states. The introduction of anisotropy does provide a small
improvement to the correlation function of high-momentum states a few
Euclidean time slices after the source.

Our results indicate that future studies of high-momentum states
should adopt this relatively cheap program of tuning the smearing
parameters to optimize isolation and overlap with the states of
interest.  We anticipate this approach will be of significant benefit
in future form factor studies.

\section{Acknowledgments}

This research was undertaken on the NCI National Facility in Canberra,
Australia, which is supported by the Australian Commonwealth
Government. This research is supported by the Australian Research
Council.


\end{document}